\documentstyle[aps,pra,twocolumn]{revtex}
\input epsf

\begin{document}
\draft
\title{Bose-Einstein condensation in a stiff TOP trap with adjustable geometry}
\author{E. Hodby, G. Hechenblaikner, O.M. Marag\`o, J. Arlt, S. Hopkins\thanks{Present address:
SCOAP, CPES, University of Sussex, Falmer, Brighton BN1 9QH} and C. J. Foot}
\address{Clarendon Laboratory, Department of Physics, University of Oxford,\\
Parks Road, Oxford, OX1 3PU, \\
United Kingdom.}
\date{28th March 2000}
\maketitle

\noindent

\begin{abstract}
We report on the realisation of a stiff magnetic trap with
independently adjustable trap frequencies, $\omega_z$ and $\omega_r$,
in the axial and radial directions respectively. This has been
achieved by applying an axial modulation to a Time-averaged Orbiting
Potential (TOP) trap. The frequency ratio of the trap, $\omega_z /
\omega_r$, can be decreased continuously from the original TOP trap
value of 2.83 down to 1.6. We have transferred a Bose-Einstein
condensate (BEC) into this trap and obtained very good agreement
between its observed anisotropic expansion and the hydrodynamic
predictions. Our method can be extended to obtain a spherical
trapping potential, which has a geometry of particular theoretical interest.
\end{abstract}
\pacs{PACS numbers: 03.75.Fi, 05.30.Jp, 32.80.Pj}

Since the first observation of BEC in a dilute alkali gas in 1995 \cite{anderson},
the spectroscopy of the excitations of these novel quantum objects has
been an important tool for understanding them. A close analogy can be
drawn between the discrete excited states of a condensate and the well defined energy levels of an atom.
Spectroscopic measurements of atomic energy levels have guided the development of the quantum
mechanical model of the atom and likewise the spectroscopy of the condensate should provide crucial
information about its nature. The first experiments focused on
 low energy collective excitations both in a pure condensate close to absolute zero \cite{jin} and at finite temperature \cite{jin2}.
These latter experiments led to a better understanding of the
interaction between the condensate and the thermal cloud
\cite{hutchinson}. The first observation of a transverse collective excitation was recently
reported in \cite{scissors}. This also gave evidence for superfluidity in a condensate.
Recent theoretical work has provided information on the behaviour of an excited condensate in
specific anisotropic trap geometries. This paper presents an adjustable magnetic trap that
allows us to access these geometries whilst retaining a stiff
trapping potential.

In the linear hydrodynamic theory of Bose-Einstein condensation,
the spectrum of low energy collective excited states depends
strongly on the trap geometry (i.e. the trap stiffnesses in all
three orthogonal directions). By changing the trap frequency in
one specific direction, the mode energies change relative to one
another and new features appear in the spectrum. For smaller
condensates, the full Gross-Pitaevskii equation including the
kinetic energy term is required. In this regime, the size of the
condensate, as well as its shape, plays an important role in
determining the excitation spectrum. If the shape of the potential
is adjusted so that the second harmonic of one mode corresponds to
the fundamental frequency of another, then the inherent
non-linearity of the system leads to coupling between the modes.
It has been predicted from the hydrodynamic limit of the
Gross-Pitaevskii equation that this behaviour should be observed
for a trap potentials with frequency ratios of 1.95, 1.51 and 0.68
\cite{dalfovo}.

Another geometry of great interest is the spherically symmetric trap,
because the excitation spectrum of the condensate is simplified by
degeneracy and becomes theoretically tractable \cite{pit}. One mechanism for the
decay of collective excitations in a condensate is Landau damping,
which is analogous to a scattering process. A condensate
excitation disappears in a collision with a thermal excitation,
resulting in a thermal excitation at a higher energy. The reduced
availability of states in a spherical trap, caused by degeneracy, has a major effect
on the Landau damping rate. Quantitative calculations of this process are currently
being carried out \cite{rusch}.

In addition to its importance for studying the collective excitations of a BEC,
the spherically symmetric harmonic trap also produces unique behaviour in a
thermal cloud of atoms. The breathing mode oscillation is completely undamped
in this geometry \cite{boltzmann}, whilst in an anisotropic trap all modes are strongly damped \cite{GO}.

There are two general types of purely magnetic trap for neutral
atoms. Ioffe-Pritchard type traps \cite{IP} use static magnetic fields to
generate a very tight radial confinement. They produce prolate
condensates with aspect ratios of $\sim 0.05$. The aspect ratio of a
condensate or thermal cloud is the ratio of the radial to axial size,
and is the inverse of the trap frequency ratio, in a harmonic
potential. The TOP trap \cite{TOP}, in its axially symmetric configuration,
combines a quadrupole field with a rotating bias field to trap an
oblate condensate, with an aspect ratio of 2.83. In this paper we
demonstrate a novel method to independently relax the axial trapping
frequency of the TOP trap. This new trap, with its adjustable aspect
ratio, has been called a zTOP trap because of the configuration of bias fields that is used.

The original attempts to vary the aspect ratio of the TOP trap made
use of the force of gravity on atoms held in a very weak magnetic
potential \cite{ensher}. This method, along with its practical difficulties, is
discussed briefly. In contrast, the zTOP trap combines an adjustable
aspect ratio with a stiff trapping potential. The experimental setup
is described and finally evidence for the successful transfer of a
condensate into the zTOP trap is presented.

The standard TOP trap combines a spherical quadrupole field of radial gradient
$B_q^{\prime}$, with a bias field, $\mathbf{B_t}$, rotating
in the horizontal xy-plane. The expression for the trap
potential usually ignores gravity. However, for heavy atoms such as
$^{87}$Rb, in a weak trap, it must be included. The full potential is given by the time-average of Eq.~\ref{pot1}, where $r_0 =
B_t / B_q^{\prime}$.

\begin{eqnarray}
\lefteqn{U(x,y,z,t) =  \mu B_q^{\prime}} \nonumber \\ &&
\left( | \left( x+r_0 \cos \omega_t t \right) {\mathbf{\hat{x}}} +
\left( y+r_0 \sin \omega_t t \right) {\mathbf{\hat{y}}} - 2z
{\mathbf{\hat{z}}} | + \frac{mgz}{\mu B_q^{\prime} } \right)
\label{pot1}
\end{eqnarray}

At small distances from the trap centre, $ r \ll r_0 $, this approximates to
a harmonic trapping potential. The trap oscillation frequencies and
frequency ratio are

\begin{eqnarray}
f_r & = & \frac{1}{2 \pi} \sqrt{\frac{\mu}{2 m}}
\frac{B_q^{\prime}}{\sqrt{B_t}}(1-\eta^2)^{1/4} (1+\eta^2)^{1/2}
\label{r} \\ f_z & = & \frac{1}{2 \pi} \sqrt{\frac{\mu}{2 m}}
\frac{B_q^{\prime}}{\sqrt{B_t}} \sqrt{8} (1-\eta^2)^{3/4} \label{z}
\\
\frac{f_z}{f_r}& = & \sqrt{\frac{8 (1-\eta^2)}{1+\eta^2}}
\label{AR}
\end{eqnarray}
where the parameter
\begin{equation}
\eta = \frac{mg}{2 \mu B_q^{\prime}} \label{nu}
\end{equation}
is the ratio of
the gravitational force on the atom to the axial magnetic force due to the
quadrupole field. Note that the trap frequency ratio depends only on $\eta$. Thus by  measuring the axial and radial trap frequencies,
as a function of $B_q^{\prime}$, we were able to accurately calibrate
the quadrupole field strength.

However, the aspect ratio only changes significantly in very weak traps, where
the gravitational force becomes important. A ten percent
decrease in the aspect ratio requires $B_q^{\prime} < 23$ G/cm. Since
$B_t$ is limited by noise to a minimum value of 2$\,$G in our apparatus, this corresponds
to a radial trapping frequency of $< 15$ Hz. Attempts were
made to selectively excite quadrupole oscillations of the BEC in this
weak gravity trap. However, in such a weak trap it was impossible not
to excite large dipole motions, making the quadrupole oscillation
impossible to observe.

The theory of the zTOP trap is analogous to that for the standard TOP trap,
except that the bias field oscillates in three rather than two dimensions.
Both are based on a spherical quadrupole field trap, with a trapping potential of
\begin{equation}
U(x,y,z) = \mu B_q^{\prime} \left| x {\mathbf{\hat{x}}}  + y {\mathbf{\hat{y}}}  -2z {\mathbf{\hat{z}}}  \right|
\label{quad}
\end{equation}
which has a restoring force in the axial direction that is
twice as strong as that in the radial direction. Now consider the effect of
applying a TOP bias field rotating at $\omega_t$ in the (radial)
xy-plane. This causes the locus of the quadrupole field to
describe a circle in the xy-plane and after time-averaging creates
an axially symmetric TOP trap. The trap stiffness is reduced in
all directions compared to the quadrupole field trap, but more
significantly in the radial direction, increasing the aspect ratio
to 2.83. Extending this argument, the trap stiffness can be
preferentially reduced in the z direction by including an
oscillating axial bias field. Our present arrangement uses an
axial modulation frequency ($\omega_a$) of $2 \, \omega_t$. Thus in
our zTOP trap, the locus of the quadrupole field follows a three
dimensional saddle shape rather than a circle. The total zTOP
magnetic field is given in Eq.~\ref{field} as the sum of its three
components - the static quadrupole field and the oscillating
radial and axial bias fields.

\begin{eqnarray}
\lefteqn{{\mathbf B}(t) =
 B'_q\left(x{\mathbf{\hat{x}}}+y{\mathbf{\hat{y}}}-2z{\mathbf{\hat{z}}}\right)+}
\nonumber \\ && B_t \left(\cos \omega_t t\ {\mathbf{\hat{x}}}+\sin
\omega_t t\ {\mathbf{\hat{y}}} \right)+B_z\cos \omega_a t\
{\mathbf{\hat{z}}} \label{field}
\end{eqnarray}

The properties of the zTOP trap are calculated from the instantaneous zTOP potential, numerically averaged over one
cycle of the bias field. This calculation gives the following constraints on
the frequency, $\omega_a$, of the axial
modulation necessary to create a stable trap.
 Firstly, to create a time-averaged potential, $\omega_a$
must be greater than the trap frequencies. Secondly $\omega_a$ must be
an exact integer multiple, n, of the radial (TOP) bias field frequency,
$\omega_t$. Our calculations showed that deviations from this condition produced asymmetry in the time-averaged potential between the x and y directions.
In the xy-plane, the trap will acquire a slightly elliptical, rotating cross-section, whilst its centre will describe a small circle.
 These slow variations in the time-averaged
potential at the beat frequency, $\omega_a$ - n$\, \omega_t$, could
 heat the cloud. It is also possible to choose $\omega_t$ as an integer multiple of $\omega_a$. However this configuration
requires the two signals to be phase locked, if the 'micromotion' of the trap described above is to be avoided.  In addition, $\omega_a$ must be
lower than the Larmor frequency, so that Majorana spin flips are not
induced.

\begin{figure}
\begin{center}\mbox{ \epsfxsize 3in\epsfbox{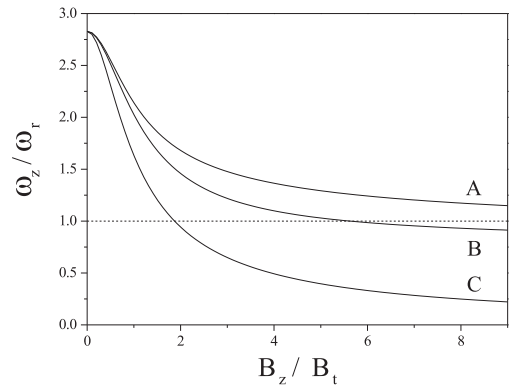}}\end{center}
\caption{Simulations of the trap frequency ratio $\omega_z / \omega_r$
as a function of the $B_z / B_t$. The solid lines correspond to
different $B_z$ waveforms, (A) fundamental only, (B) fundamental +
third harmonic component of a square wave, (C) square wave. The dotted line indicates the
value of $B_z / B_t$ required for a spherically symmetric
trap in each case.}\label{sim}
\end {figure}

The extent of the trap relaxation in the axial direction depends
on the waveform and amplitude, $B_z$, of the axial modulation, but
not on its frequency. The axial modulation causes the locus of the quadrupole field to
oscillate in the vertical direction. A waveform that causes it to spend more time per cycle at the extremes
of this motion results in a more relaxed trap in the axial direction. Thus an ideal square wave produces the
weakest trap, for a given maximum $B_z$ field. Removing the higher
harmonics in the axial modulation stiffens the trap as shown in
Fig.\ref{sim}. Note that significant additional relaxation, with respect to single frequency modulation,
is achieved by adding only the third harmonic component of a square wave to the
fundamental. In this case a spherical trap is produced for $B_z
/ B_t = 5.5$. Since $B_t$ is typically 2$\,$G, this is a readily
achievable experimental condition. The axial trapping frequency is also reduced by increasing the modulation
amplitude, $B_z$. Thus we are able to vary the aspect ratio of the
trap during the course of an experiment by controlling the peak
voltage across the axial bias coils.

The apparatus that we use to create $^{87}$Rb condensates is
described in detail elsewhere \cite{rotation}. In summary, we use a differentially
pumped double MOT system (which incorporates a pyramidal MOT \cite{pyramid}) to load
a TOP trap with $\sim 2 \times 10^8$ atoms. Evaporative cooling
proceeds via both Majorana spin flip and radio frequency cutting to
achieve condensates with $\sim 2 \times 10^4$ atoms.

To generate the axial bias field, two Helmholtz coils of 35 turns
each, have been added above and below the experimental cell. A 2 turn
pickup coil is also present to monitor the applied field. We
currently use an axial modulation at 14 kHz (2$\omega_t$). The
amplitude of this signal is computer controlled and fed to an audio
amplifier. A transformer is used to match the output of the amplifier
to the coils. Fine adjustments are made to the frequency until the x
and z pickup coils produce a clean Lissajous figure on the
oscilloscope, indicating that the axial oscillation is at exactly
twice the frequency of the radial bias field. This frequency
relationship is stable to within $\sim 1$ Hz over the day.

\begin{figure}
\begin{center}\mbox{ \epsfxsize 3in\epsfbox{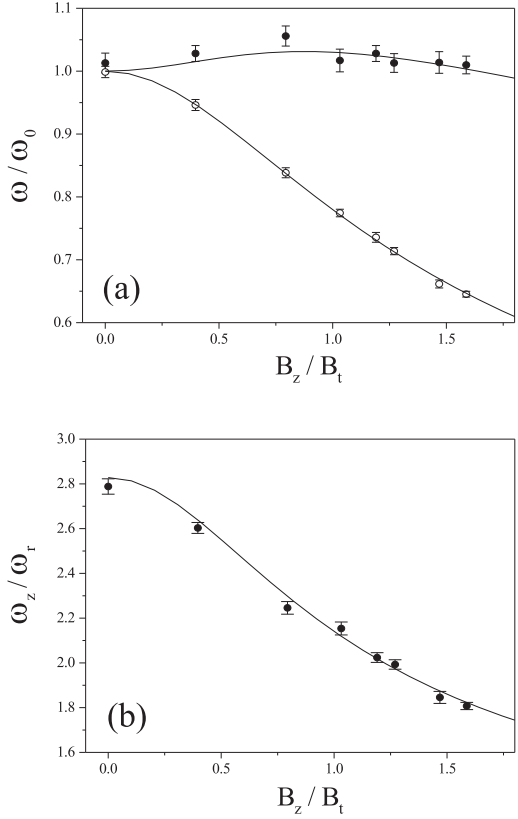}}\end{center}
\caption{Experimental values for the normalised trap frequencies and aspect ratio as a function
of $B_z / B_t$, with the theoretical curve that was used to calibrate the
zTOP trap. (a) shows how the  normalised individual trap frequencies, $\omega_r /
\omega_{r0}$ (solid circles) and $\omega_z / \omega_{z0}$ (open
circles), depend on $B_z / B_t$. $\omega_{r0}$ and $\omega_{z0}$ are the
radial and axial trap frequencies when $B_z = 0$. (b) shows the
aspect ratio of the trap as a function of $B_z / B_t$. For all points, $B_t
= 2G$.}\label{dip}
\end {figure}

The trap frequencies were calibrated with respect to
the voltage supplied across the axial bias field coils. A small, cold
thermal cloud was collected in a 'displaced' zTOP trap. The
additional static magnetic fields causing the displacement were
suddenly switched off, exciting orthogonal dipole oscillations at the
two trap frequencies. The radial and axial positions of the cloud
were recorded as a function of time, using absorption imaging, and
fitted with sine waves to obtain the trap frequencies. Figure \ref{dip}
shows the data for the normalised trap frequencies and the trap frequency ratio, for a
range of axial bias field amplitudes.

Theoretical values for the normalised trap frequencies, as a function of $B_z / B_t$, were found by numerically integrating the magnetic potential
energy of an atom over one cycle of the bias field. These values were accurately described by an eighth order
polynomial. Two free fitting parameters were used to fit this polynomial to the experimental data.
One parameter determined the trap frequencies without any axial bias field and the other was a constant,
relating the voltage across the axial bias coils to the actual value of $B_z/B_t$ that was produced.
Fitting this expression to the axial trap frequency data gave values for both
fitting parameters and their respective errors.
The solid lines in Fig.\ref{dip} show the fitted zTOP trap calibration curves.

The condensate is initially produced by a sequence of computer
controlled evaporative cooling ramps in a standard TOP trap, with
final trap frequencies $\omega_r = 126$ Hz and $\omega_z = 356$ Hz
$(= \sqrt{8} \, \omega_r)$.
 The amplitude of the oscillating axial field is then linearly ramped from zero to its final value over 0.5s.
This adiabatically transfers the condensate into the zTOP trap.

\begin{figure}
\begin{center}\mbox{ \epsfxsize 3in\epsfbox{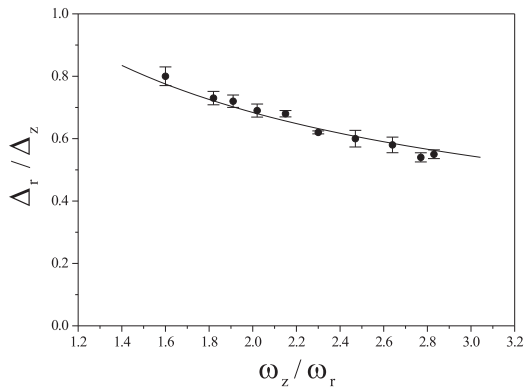}}\end{center}
\caption{The aspect ratio of a BEC, ($\Delta
r / \Delta z$), after 12ms time-of-flight, as a function of the zTOP trap frequency ratio
($\omega_z / \omega_r$).
 The solid line shows the solution of the hydrodynamic equations as a function of trap geometry (no free parameters). Each data point is the mean of six shots and the error bar shows the standard deviation of the mean. }\label{cd}
\end {figure}

Figure \ref{cd} shows the aspect ratio of a BEC, 12ms after release
from zTOP traps with a range of frequency ratios. Changing the
frequency ratio of the trap changes the distribution of the self
energy of the condensate among the three orthogonal directions and
hence the form of the free expansion of the condensate. The aspect
ratio of the expanding BEC has been predicted as a function of trap
geometry using the hydrodynamic equations \cite{castin} and is shown as a solid line
in Fig.~\ref{cd}. This theoretical curve contains no free parameters.
The good agreement between theory and experiment
confirms two facts. Firstly that the condensed state survived adiabatic
transfer to the zTOP trap and secondly that the trap potential
experienced by the condensate has been modified as predicted.

In conclusion, we have modified our TOP trap to create a stiff zTOP trap with an
adjustable geometry. Trap frequency ratios between 2.83 and 1.6 have
been demonstrated by applying a magnetic field in the axial
direction, oscillating at a single frequency. The addition of higher
harmonics to the axial bias field modulation will make a spherical
trap accessible. This prediction, combined with the successful
transfer of BECs into the zTOP trap, as demonstrated in this paper,
promises to open up the field of degenerate condensate excitations to
experimental reasearch.

This work was supported by the EPSRC and the TMR program (No. ERB
FMRX-CT96-0002). O.M. Marag\`{o} acknowledges the support of a
Marie Curie Fellowship, TMR program (No. ERB FMBI-CT98-3077).

\begin{thebibliography}{99}


%%%%%%%%%%   The first BECs
\bibitem{anderson} M.H. Anderson {\it et al.}, Science {\bf 269}, 198 (1995);
                   K.B. Davis {\it et al.}, Phys. Rev. Lett. {\bf 75}, 3969 (1995).
For a review see Bose-Einstein Condensation in Atomic Gases,
Proceedings of the International School of Physics ``Enrico
Fermi'', edited by M. Inguscio, S. Stringari and C.E. Wieman,
(IOS Press, Amsterdam, 1999).

%%%%%%%%%    JILA and MIT experiments on collective excitations
\bibitem{jin}
          D. S. Jin, J.R. Ensher, M.R. Matthews, C.E. Wieman and E. A. Cornell, Phys. Rev. Lett. {\bf 77}, 420 (1996); \\
M.O. Mewes, M.R. Andrews, N.J. van Druten, D.M. Kurn, C.G. Townsend and W. Ketterle, Phys. Rev. Lett. {\bf 77}, 988 (1996)

%%%%%%%%% Experiment on temperature dependence of collective excitations
\bibitem{jin2}
           D.S. Jin, M.R. Matthews,J.R. Ensher,  C.E. Wieman and E. A. Cornell,
           Phys. Rev. Lett. {\bf 78}, 764 (1997); \\
D.M. Stamper-Kurn, H-J Meisner, S. Inouye, M.R. Andrews and W. Ketterle, Phys. Rev. Lett. {\bf 81}, 500 (1998)

\bibitem{hutchinson}
           D.A.W. Hutchinson, R.J. Dodd, and K. Burnett, Phys. Rev. Lett. {\bf 81},
           2198 (1998)

%%%%%%%%%% Experimental Superfluidity
\bibitem{scissors} O.M. Marag\`{o}, S.A. Hopkins, J. Arlt, E. Hodby, G. Hechenblaikner and C.J. Foot, Phys. Rev. Lett. {\bf 84}, 2056 (2000)


%%%%%%%%Non-linear Papers
\bibitem{dalfovo}
          F. Dalfovo, C. Minniti, and L.P. Pitaevskii,
          Phys. Rev. A {\bf 56}, 4855 (1997);\\
G. Hechenblaikner, O.M. Marag\`{o}, E. Hodby, J. Arlt, S. Hopkins and C.J. Foot, cond-mat/0003495

%%%%%%% Spherical Stuff
\bibitem{pit}
         H. Guilleumes and L. P. Pitaevskii, Phys. Rev. A {\bf 61}, 013602 (2000)
\bibitem{rusch}M. Rusch, private communication
\bibitem{boltzmann} L. Boltzmann in {\it {Wissenschaftliche Abhandlungen,}} edited by F. Hasenori (Barth, Leipzig, 1909), Vol.2, p.83
\bibitem{GO}D. Gu\'{e}ry-Odelin, F. Zambelli, J. Dalibard and S. Stringari, Phys. Rev. A {\bf 60}, 4851 (1999)
%%%%%% Magnetic traps
\bibitem{IP} D. Pritchard, Phys. Rev. Lett. {\bf 51}, 1336 (1983)
\bibitem{TOP} W. Petrich, M. Anderson, J. Ensher and E.A. Cornell, Phys. Rev. Lett. {\bf 74}, 3352 (1995)
\bibitem{ensher} J.R. Ensher, PhD thesis, University of Colorado (1998)

\bibitem{rotation} J. Arlt, O.M. Marag\`{o}, E. Hodby, S.A. Hopkins, G. Hechenblaikner and C.J. Foot, J. Phys. B, {\bf 32} 5861 (1999)
\bibitem{pyramid} J. Arlt, O. Marag\`{o}, S. Webster, S. Hopkins and C.J. Foot, Opt. Comm. {\bf 157}, 303 (1998)
\bibitem{castin} Y. Castin and R. Dum, Phys. Rev. Lett. {\bf 77} 5315 (1996)

\end{thebibliography}
\end{document}